\documentstyle[aps,prb,epsf,floats,twocolumn]{revtex}
\begin{document}
\draft
\title{Dephasing in a quantum pump}
\author{J.N.H.J. Cremers,$^a$ P.W. Brouwer$^b$}
\address{$^{a}$Lyman Laboratory of Physics, Harvard University, Cambridge MA 
02138\\$^{b}$Laboratory of Atomic and Solid State Physics,
Cornell University, Ithaca, NY 14853-2501\\ {\rm (\today)}
\medskip  ~~\\\parbox{14cm}{\rm
We study how dephasing affects the
distribution of the dc current pumped through a
chaotic quantum dot. We introduce dephasing by the addition of 
a voltage probe to the quantum dot, treating both the
case of controlled dephasing (when the voltage probe is coupled
to the dot via a ballistic point contact with a conductance that
is increased stepwise) and
intrinsic dephasing (for which the voltage probe serves as a
phenomelogical model). While dephasing
eventually suppresses the dc current through the dot,
we also find that, for a quantum dot with single-mode point contacts,
a small amount of dephasing actually decreases the likelihood of a 
zero pumped current.
\pacs{PACS numbers: 72.10.Bg, 73.23.-b}}}

\maketitle

\section{Introduction}\label{intro}

Quantum mechanical interference is a key player determining
sample-specific
properties of small metal or semiconductor particles. For closed
systems, quantum interference fixes the precise
positions of energy levels and
the microscopic details of wavefunctions; for open systems it governs,
e.g., the fluctuations of the conductance.\cite{LesHouches,Kouwenhoven}
In all these cases,
interference provides a (sometimes large) correction added to a
nonzero background. Such a background is absent in a so-called
``quantum electron pump'', which has received considerable 
experimental\cite{switkes} and 
theoretical\cite{SZB,B1,ZSA,AG,SAA,Avron,VAA,AAK,polianski} 
attention recently. 
This requirement of phase coherence for the observation of a 
pumped current makes a quantum pump a sensitive instrument to 
study quantum interference in quantum dots.

In the experimental proposal of Ref.\ \onlinecite{switkes}, a 
quantum electron pump is made of a semiconductor 
quantum dot connected to two electron reservoirs. A dc current is 
generated by periodically 
varying two gate voltages that characterize shape of the dot.
(Alternatively, one may pump electrons by periodic variation of the 
Fermi energy, magnetic flux, etc.) The electron pump
is called ``quantum'', because the variation pertains to the quantum
interference only, not to the classical dynamics of the dot: In a
trajectory-based picture, the variations affect phases of the
trajectories, not their weights. For this reason, this 
``quantum'' electron pump is fundamentally different from
electron pumps or turnstiles that 
rely on Coulomb charging physics and do not
require phase coherence.\cite{KJVHF,PLUED} In the quantum pump,
the precise value of the pumped current and its direction depend
on microscopic details and vary from sample to sample. Therefore,
one needs to address 
the probability distribution $P(I)$
of pumped currents for an ensemble of
quantum pumps. Theoretical work has focused on electron pumps in a
chaotic quantum dot, for which this distribution can be calculated
using random matrix theory.\cite{B1,ZSA,SAA,VAA} For a pump with
single-channel point contacts, it was found that $P(I)$ has a
cusp at $I=0$ and algebraic tails,\cite{B1} while for many-channel point
contacts, $P(I)$ is Gaussian.\cite{B1,ZSA,SAA}

In this paper, we study how the probability
distribution of the pumped current is affected if phase coherence
in the quantum pump is gradually destroyed. As in Refs.\ 
\onlinecite{switkes,B1,ZSA,AG,SAA,Avron,VAA}, 
we consider the case of a quantum pump
that consists of a chaotic quantum dot.
A controlled way to destroy phase coherence is to
couple the dot to a voltage probe.\cite{but3}
While not drawing a net current, the voltage
probe absorbs and reinjects electrons without any phase-relationship, 
thereby destroying phase coherence. The dephasing rate $\gamma_{\phi}$
can be tuned by varying the conductance
of the point contact connecting
the quantum dot and the voltage probe.
While a voltage probe can be
used as an external source to controllably increase the dephasing
rate in the
quantum pump, it can also be used as a model for intrinsic 
dephasing processes in the pump itself, such as electron electron 
interactions or two-level systems in the dot.\cite{but3,BM1,Brouwer}
A similar approach was taken in a recent paper by Moskalets and 
B\"{u}ttiker, who studied the effect of a voltage probe on the
current pumped through an electron pump consisting
of a one-dimensional wire with two tunable tunnel barriers of oscillating 
strength.\cite{MB} However, unlike the quantum dot geometry studied
here, the pump of Ref.\ \onlinecite{MB} can also operate as an
electron pump in the absence of phase coherence. The variance of
the pumped current in the case of many-channel point contacts 
was previously obtained by Shutenko {\em et al.} using a different
model.\cite{SAA}

In Sec.\ \ref{sec2} we use the scattering approach to derive
a formula for the pumped current in the presence of the voltage
probe. In Sec.\ \ref{sec3} we then
study the distribution of the pumped current for the
cases of controlled dephasing and intrinsic dephasing. We focus
on the cases of a quantum dot with single channel point contacts
and with multi-channel point contacts. We conclude in 
Sec.\ \ref{sec4}.

\section{pumped current in the presence of dephasing}
\label{sec2}

\begin{figure}
  \epsfxsize=0.9\hsize
  \hspace{0.04\hsize}
  \epsffile{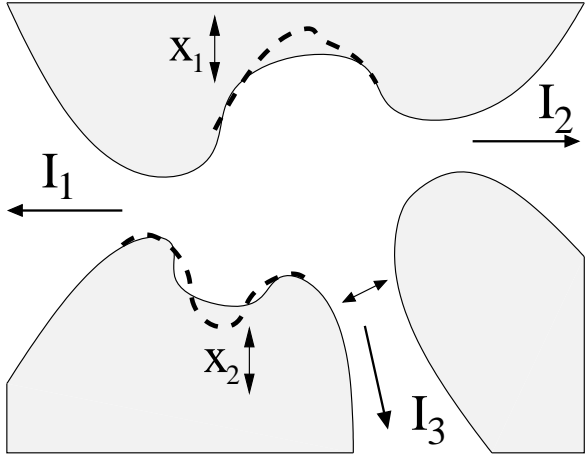}\\

\caption{ \label{fig:1} A quantum dot with two parameters $X_{1}$ and $X_{2}$
which describe a deformation of its shape. As $X_{1}$ and $X_{2}$ are varied
periodically a dc current $I=I_1=-I_2$ flows from the right to the left
reservoir. The lead on the bottom, through which no current 
flows ($I_3=0$), has a variable number of channels $N_{3}$.}
\end{figure}

A schematic picture of the system is shown in Fig.\ref{fig:1}.
It consists of a quantum dot connected by ballistic point contacts to
three electron reservoirs, labeled $1$, $2$, and $3$. The point contacts
allow $N_1$, $N_2$ and $N_3$ propagating channels at the Fermi level,
respectively. The total number of channels in all point contacts
will be denoted as $N = N_1 + N_2 + N_3$. The reservoirs
$1$ and $2$ are held at the same voltage $V_1 = V_2 = 0$,
while reservoir $3$
serves as a voltage probe: the time-dependent
voltage $V_3(t)$ is adjusted such
that the current $I_3(t) = 0$ at all times.

Two external
parameters $X_1(t)$ and $X_2(t)$ that determine the shape (or other
characteristics) of the quantum dot are varied periodically
with frequency $\omega$. As a result of this periodic variation,
a dc current $I = I_1 = - I_2$
will flow between reservoirs $1$ and $2$. We will
now evaluate this pumped current in the presence of the voltage
probe. Starting point of our evaluation is the relation\cite{B1}
between the pumped current and the emissivity $e dn(m)/dX_j$, the
charge that leaves the dot through contact $m=1,2$ as the parameter
$X_j$ is varied by an amount $dX_j$,
\begin{equation}
  I =
  {\omega e \over 2 \pi}
  \int_A dX_1 dX_2 \left[ {\partial \over \partial X_1}
  \left( {dn(1) \over dX_2} \right) -
  {\partial \over \partial X_2}
  \left( {dn(1) \over dX_1} \right) \right].
  \label{eq:Ipump}
\end{equation}
Here $A$ is the area in $(X_1,X_2)$-space enclosed by the parameters
$[X_1(t),X_2(t)]$ in one cycle. Equation (\ref{eq:Ipump}) is valid
for the adiabatic regime, $\omega \tau_d \ll 1$, where $\tau_d
\sim \hbar/N \Delta$ is the dwell time of the quantum dot,
$\Delta$ being the mean single-particle level spacing.

In the absence of the voltage probe, and for ballistic point
contacts between the dot and reservoirs $1$ and $2$, so that
effects of charge quantization in the dot can be ignored,\cite{ABG}
the emissivity $e dn(m)/dX_j$
is related to the $N \times N$ scattering matrix $S$ of the dot
as\cite{but2}
\begin{eqnarray}
  {dn(m) \over dX} &=&
  {1 \over 2 \pi} \sum_{\beta}
  \sum_{\alpha \in m} \mbox{Im}\, {\partial S_{\alpha \beta} \over
    \partial X} S_{\alpha \beta}^{*}.  \label{eq:emis}
\end{eqnarray}
With voltage probe, charge can either leave the quantum dot
directly through the contacts $1$ or $2$, or
via inelastic scattering in reservoir $3$. Hence,
we can write the emissivity
$dn(m)/dX$ as a sum of an elastic and an inelastic contribution,
\begin{equation}
  {dn(m) \over dX} =
  \left(dn(m) \over dX \right)_{\rm el} +
  \left(dn(m) \over dX \right)_{\rm in}, \label{eq:emiselinel}
\end{equation}
where
the elastic contribution
$(dn(m)/dX)_{\rm el}$ is still given by Eq.\ (\ref{eq:emis}) and
the inelastic contribution reads
\begin{eqnarray}
  \left({dn(m) \over dX} \right)_{\rm in} &=&
  {G_{m3} \over 2 \pi(G_{13} + G_{23})}
  \mbox{Im}\, \sum_{\beta}
  \sum_{\alpha \in 3} {\partial S_{\alpha \beta} \over
    \partial X} S_{\alpha \beta}^{*}.  \label{eq:emisinel}
\end{eqnarray}
Here
$$
  G_{ij} = \sum_{\alpha \in i} \sum_{\beta \in j}
  |S_{\alpha \beta}|^2
$$
is an element of the $3 \times 3$ condutance matrix of the dot.
Substitution of Eqs.\ (\ref{eq:emis})--(\ref{eq:emisinel}) into
Eq.\ (\ref{eq:Ipump}) then yields the current formula
\begin{mathletters} \label{eq:Ipumptotal}
\begin{eqnarray}
I &=&
\frac{\omega e}{2\pi }
\int_{A}dX_{1}dX_{2} \left( i_{\rm dir} + i_{\rm rect}\right),\label{eq:pump}\\
i_{\rm dir} &=& {G_{23} \over \pi(G_{13} + G_{23})}
  \mbox{Im}\, \sum_{\beta} \sum_{\alpha\in 1}
  {\partial S_{\alpha \beta} \over \partial X_2}
  {\partial S_{\alpha \beta}^* \over \partial X_1}
  \nonumber \\ && \mbox{} -
  {G_{13} \over \pi(G_{13} + G_{23})}
  \mbox{Im}\, \sum_{\beta} \sum_{\alpha\in 2}
  {\partial S_{\alpha \beta} \over \partial X_2}
  {\partial S_{\alpha \beta}^* \over \partial X_1},\\ \label{eq:pumpa}
i_{\rm rect} &=& {1 \over 4 \pi}
  \mbox{Im}\, \sum_{\beta} \sum_{\alpha\in 3}
   S_{\alpha \beta}^{*} {\partial S_{\alpha \beta} \over \partial X_2}
  {\partial \over \partial X_1} {G_{13} - G_{23}\over G_{13} + G_{23}}
  \nonumber \\ && \mbox{} -
  {1 \over 4 \pi}
  \mbox{Im}\, \sum_{\beta} \sum_{\alpha\in 3}
  S_{\alpha \beta}^{*} {\partial S_{\alpha \beta} \over \partial X_1}
  {\partial \over \partial X_2} {G_{13} - G_{23} \over G_{13} + G_{23}}.
\end{eqnarray}
\end{mathletters}
The first contribution $I_{\rm dir}$ represents the charge that exits
the dot either directly or indirectly via the reservoir, while the
second contribution $I_{\rm rect}$ is an additional contribution
to the dc current that arises from rectification of the voltage
$V_3(t)$.\cite{MB} One verifies that in the limit where the third
reservoir is decoupled from the quantum dot, $I_{\rm rect}$ 
vanishes, while the expression for $I_{\rm dir}$ approaches that
for current in the absence of dephasing (see Ref.\
\onlinecite{B1}).

In the next section we proceed to calculate the statistical distribution
of the current $I$ for an ensemble of chaotic quantum dots. We
restrict ourselves to the case of small pumping amplitudes $X_1 =
\delta X_1 \sin(\omega t)$, $X_2 = \delta X_2 \sin(\omega t + \phi)$,
for which the current $I$ is bilinear in $\delta X_1$ and $\delta
X_2$,
\begin{equation}
  I = {1 \over 2} e \omega i \sin \phi \delta X_1 \delta X_2,
  \ \
  i = i_{\rm dir} + i_{\rm rect}.
\end{equation}
We will
consider both the case of ``controlled dephasing'', corresponding
to a real voltage probe coupled to the quantum dot via a ballistic
point contact, and that of intrinsic dephasing, which is modeled
by a voltage probe coupled to the quantum dot via a wide tunneling
contact.\cite{Brouwer}

In the absence of dephasing, the pumped current is not symmetric
under reversal of a magnetic field through the quantum
dot.\cite{SAA,AAK} One
easily verifies that in the presence of dephasing the current $I$ 
remains asymmetric under reversal of the magnetic field, even in
the presence of strong dephasing ($N_3$ large).

\section{Distribution of the pumped current in the presence of dephasing}\label{sec3}

For an ensemble of chaotic quantum dots the distribution of the scattering
matrix
and its derivatives
is known.\cite{B3} It is most conveniently expressed through the
parameterization
\begin{eqnarray}
  S = U U',\ \ {\partial S \over \partial X_j} = U Q_j U',
  \label{eq:param}
\end{eqnarray}
where $U$ and $U'$ are $N \times N$ unitary matrices and $Q_j$ is an
$N \times N$ hermitian matrix. In the presence of time-reversal
symmetry, $U' = U^{\rm T}$ and $Q_j = Q_j^{\rm T}$.
We also introduce the matrix $Q_{\varepsilon}$ that parameterizes
the energy-derivative $\partial S/\partial \varepsilon$,
$$
  {\partial S \over \partial \varepsilon} = {2 \pi \over \Delta}
  U Q_{\varepsilon} U'.
$$
For ballistic point contacts to reservoirs $1$, $2$, and $3$, the matrices
$U$ and $U'$ are uniformly distributed in the unitary group,
independently of $Q_1$, $Q_2$, and $Q_{\varepsilon}$, while
the joint distribution of the matrices $Q_1$, $Q_2$, and $Q_{\varepsilon}$
is given by
\begin{eqnarray}
  P&\propto&
  (\det Q_\varepsilon)^{-N/2 - 2(\beta N+2-\beta)}\,
  \Theta(Q)\label{eq:dist}
  \\ && \mbox{} \times
  \exp\left[{-{\beta \over 2}\,\mbox{tr}\,
    \left(Q_\varepsilon^{-1} +
    {1 \over 8}\sum_{j=1}^{2}(Q_\varepsilon^{-1}
    Q_{j}^{\vphantom{1}})^2\right)}\right], \nonumber
\end{eqnarray}
where $\Theta(Q)=1$ if all eigenvalues of $Q$ are positive and
$\Theta(Q)=0$ otherwise and $\beta = 1$ ($2$) in the presence
(absence) of time-reversal symmetry.
Note that at fixed $Q_\varepsilon$ the distribution of $Q_1$
and $Q_2$ is Gaussian with a width set by $Q_\varepsilon$.

We will now study the distribution of the pumped dc current for
controlled dephasing and instrinsic dephasing. In each case we will first study
the variance and then the full distribution. 

\subsection{Controlled Dephasing}\label{controlled}

The effect of dephasing on the distribution of the pumped current
can be studied in a controlled setting by increasing the number $N_3$
of open channels in the voltage probe one by one. The third reservoir
serves as a true voltage probe if the charge relaxation rate of
that reservoir is much larger than the frequency $\omega$, so that
the voltage $V_3(t)$ can
adjust essentially instantaneously to balance any current $I_3(t)$
flowing into or out of that reservoir as a result of the pumping
action on the quantum dot.

\begin{figure}
  \epsfxsize=0.9\hsize
  \hspace{0.04\hsize}
  \epsffile{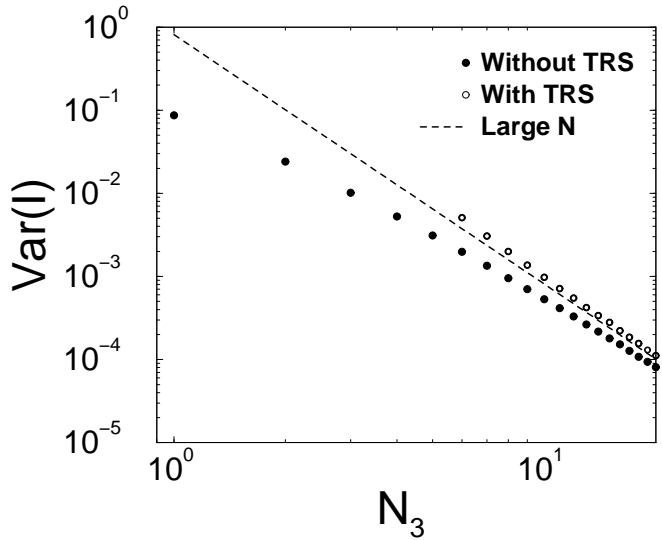} \bigskip \\

\caption{ \label{fig:2} Variance of the pumped current with one open 
channel in each of the current carrying leads ($N_1=N_2=1$)
as a function of the number of open channels in the third lead ($N_3$).
The presence (absence) of time-reversal symmetry is denoted by open
circles (closed circles). The dashed line is the variance in the asymptotic
limit $N_1=N_2=1,\,N_3\to\infty$. The variance in the presence of time-
reversal symmetry is infinite for $N_3<6$. }
\end{figure}

A voltage probe connected to the dot via a ballistic point contact
with $N_3$ channels gives rise to a dephasing rate
\begin{equation}
  \gamma_{\phi} = {1\over \tau_{\phi}} =
  N_3 \Delta/h.
  \label{eq:rate}
\end{equation}
We now fix $N_1$ and $N_2$ and calculate the distribution of
the pumped current as a function of $N_3$. We consider the cases
$N_1 = N_2 = 1$ of single channel current-carrying leads and
$N_1, N_2 \gg 1$ of many-channel current-carrying leads. In the
former case, we have calculated the first two moments of the
distribution $P(i)$ of the dimensionless current $i$
analytically, using the technique of Ref.\
\onlinecite{B4} to perform the integrations over the matrices
$U$ and $U'$. The average current $\langle i \rangle$ was thus
found to be zero, while the variance $\langle i^2 \rangle$
decreases with increasing $N_3$; $\langle i^2 \rangle$ is
divergent for $N_3 = 0$ for $\beta=2$ and for $N_3 < 6$ for
$\beta=1$. The results of this calculation
are shown in Fig.\ \ref{fig:2}. For large $N_3$, we find
\begin{equation}
\left\langle i^{2}\right\rangle=
\left\langle i_{\rm dir}^2 \right\rangle
= {16 N_1 N_2 \over \pi^2 (N_1+N_2)(N_1+N_2+N_3)^3}.
\label{eq:asym}
\end{equation}
Details of the calculation can
be found in Appendix \ref{appA}. [Equation (\ref{eq:asym}) is
valid for arbitrary $N_1$ and $N_2$, as long as $N_3 \gg 1$.] 
For $N_3 \gg 1$, the contribution of the rectification current
$i_{\rm rect}$ to $\langle i^2 \rangle$ is proportional to
$N_3^{-4}$ and hence negligible. (A
similar conclusion for the rectification contribution was reached
in Ref.\ \onlinecite{polianski} for the comparison of measurements
of pumped current and pumped voltage.)
As shown in Fig.\ \ref{fig:2},
in general, $\langle i^2 \rangle$
is larger with time-reversal symmetry than without, though the 
difference between the two variances vanishes
for large dephasing rates.

\begin{figure}

  \epsfxsize=0.9\hsize
  \hspace{0.03\hsize}
  \epsffile{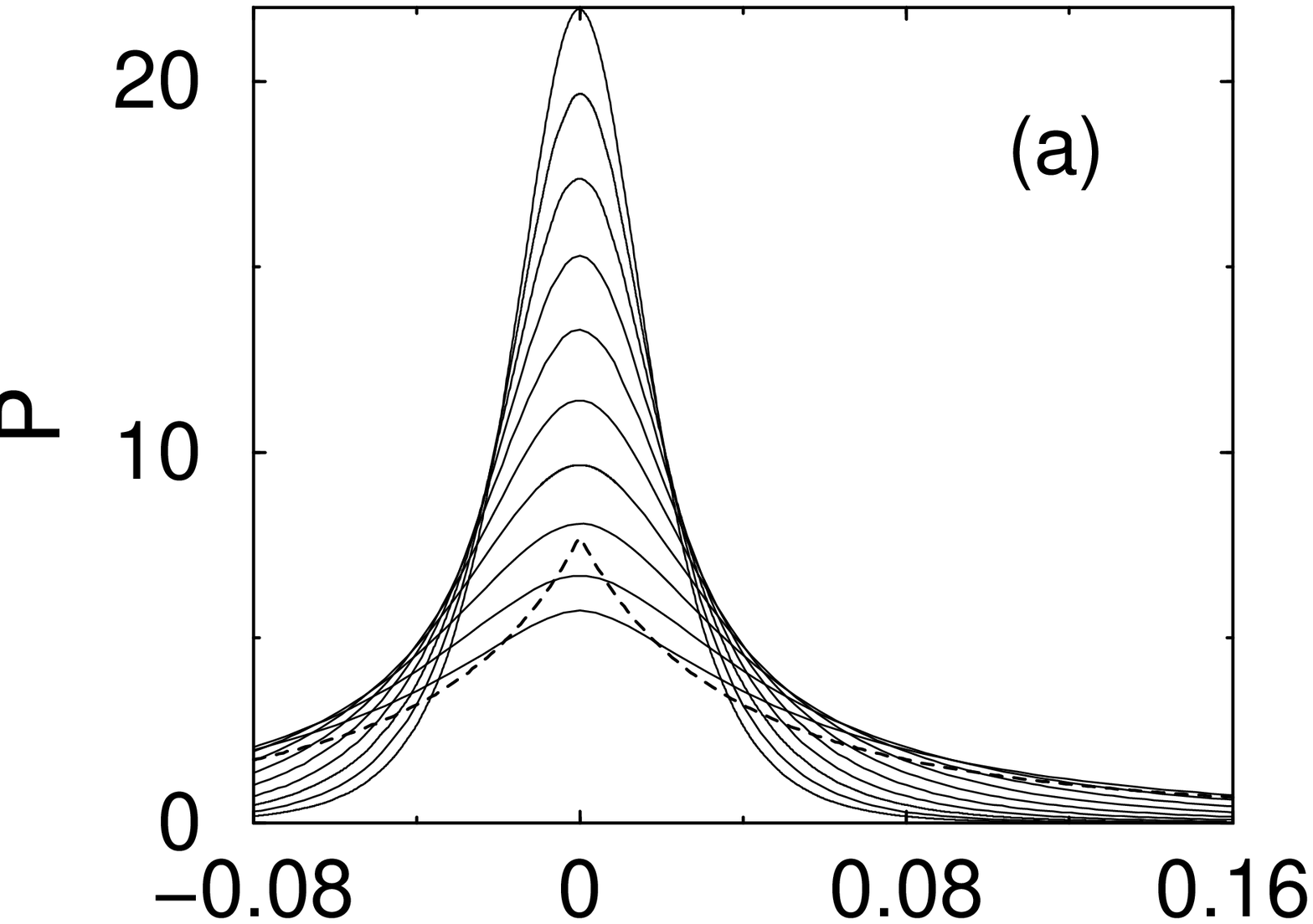}\\

  \epsfxsize=0.9\hsize
  \hspace{0.03\hsize}
  \epsffile{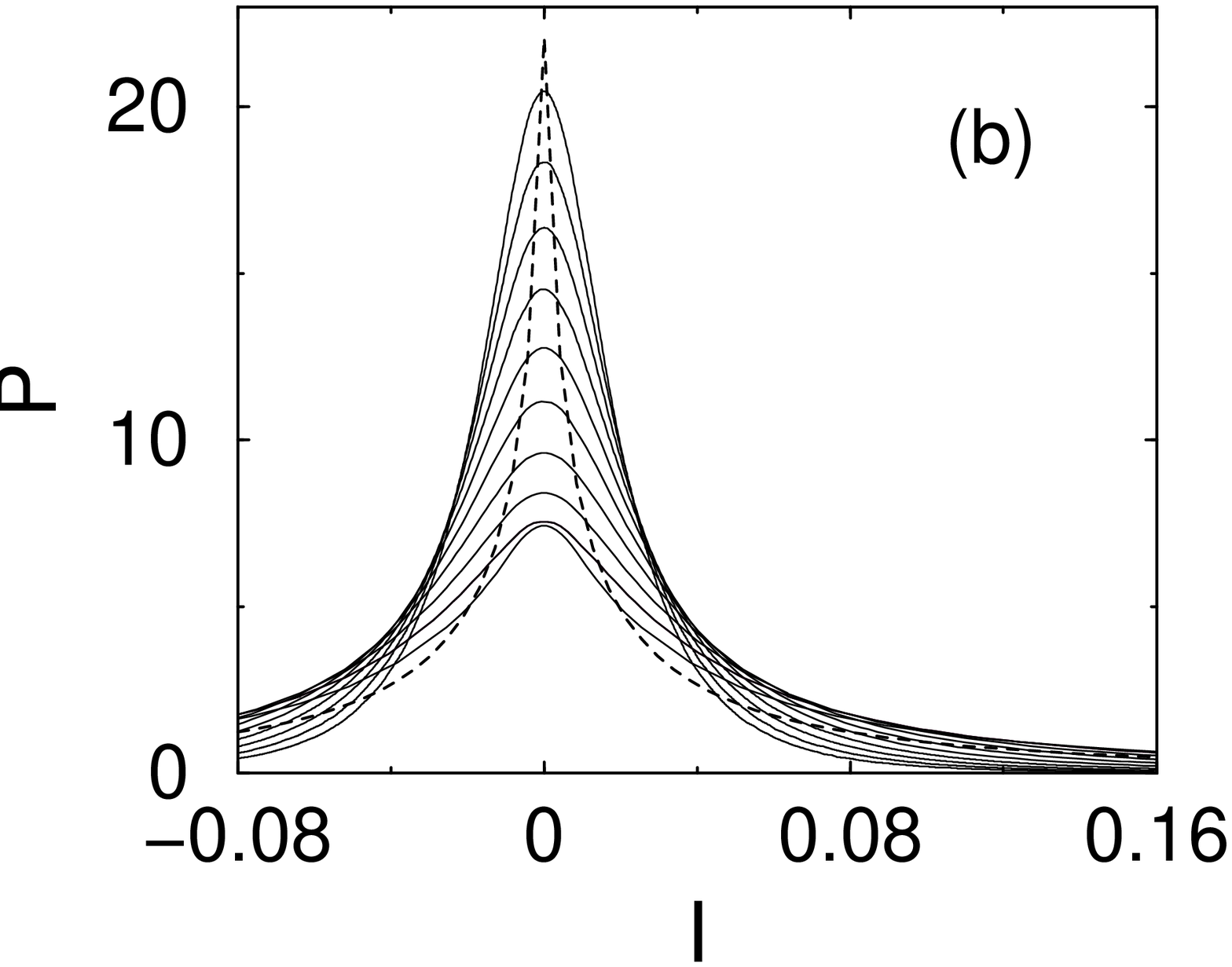}\\

\caption{ \label{fig:3} The probability distribution of the pumped current
without (a) and with (b) time-reversal symmetry, for the case $N_1 = N_2 = 1$
of single-channel point contacts in the current-carrying leads and as
a function of the number of channels $N_3$ in the voltage probe.
The dashed line shows the
distribution without dephasing ($N_3 = 0$).
The solid lines show the distribution with
$N_3=1,$ (lowest curve at $i=0$) 
$2,\, ...\, ,\,N_3=10$ (highest curve). }
\end{figure}

We obtained the full distribution $P(i)$ for $N_1 = N_2 = 1$
using Monte-Carlo integration with the distribution (\ref{eq:dist}).
The result is shown in
Fig.\ \ref{fig:3} for various values of $N_3$. Details of the
numerical procedure are outlined in App.\ \ref{appB}. Note that
the distributions are symmetric around $i=0$. In the absence of
dephasing ($N_3 = 0$),
$P(i)$ has a cusp at $i=0$, which is smoothed out for
$N_3 > 0$. The distribution $P(i)$ has algebraic tails with powers that
increase with increasing dephasing rate.
In the presence of
time-reversal symmetry the tails fall
off with a power smaller than $3$ as long as $N_3 < 6$, corresponding
to a divergent second moment $\langle i^2 \rangle$ (cf.\ Fig.\
\ref{fig:2}).
Interestingly, while dephasing decreases the second moment $\langle
i^2 \rangle$ for single-channel point contacts $N_1 = N_2 = 1$, a
small amount of dephasing also decreases the probability $P(i=0)$ of
finding no pumped current at all. This can be
understood physically, since the cusp in the distribution $P(i)$ at
zero pumped current arises from a the occurence of destructive 
interference, which is as much suppressed by
dephasing as the constructive interference responsible for large values
of $i$ and the tails of the distribution.
For $N_3 \gg 1$, the distribution $P(i)$ approaches a Gaussian.

For multichannel point contacts in the current-carrying leads,
$N_1, N_2 \gg 1$, the distribution $P(i)$ is a Gaussian centered
around $i=0$. To calculate the second moment $\langle i^2 \rangle$,
we parameterize the scattering matrix $S$ and its derivatives as
in Eq.\ (\ref{eq:param}),
perform the Gaussian integration over $Q_1$ and $Q_2$ with
the distribution (\ref{eq:dist}) at fixed $Q_{\varepsilon}$,
integrate over the unitary matrices $U$ and $U'$ using the diagrammatic
technique of Ref.\ \onlinecite{B4} and finally integrate over the
eigenvalues $\tau_j$, $j=1,\ldots,N$, of $Q_{\varepsilon}$, see
appendix \ref{appA} for details. The result is given by Eq.\
(\ref{eq:asym}) above.

\subsection{Intrinsic dephasing}

The voltage probe can also be used as a model for intrinsic dephasing
in the quantum dot. Sources of dephasing may be, e.g., electron-electron
interactions or interactions with an external bath of photons and/or
phonons. In such a case the source of decoherence is delocalized
throughout the dot. This situation is well modeled by a voltage probe
with a tunnel barrier with transmission probability $\Gamma_3 \ll 1$
and many channels $N_3 \gg 1$, such that the product
\begin{equation}
  N_3 \Gamma_3
  =  {h \gamma_{\phi} \over \Delta}
\end{equation}
is kept fixed.\cite{Brouwer}

In order to find the distribution of the pumped current in this case,
we need the distribution of the scattering matrix $S$ and its
derivatives when the contact to the third reservoir contains a
tunnel barrier. This problem can be solved by a statistical mapping
which connects the scattering matrix $S$ to a scattering matrix
$S_0$ that is taken from an ensemble appropriate for a quantum dot
with ballistic point contacts,\cite{MelloSeligman,B5}
\begin{equation}
  S = \sqrt{1 - \Gamma} - \sqrt{\Gamma} {1 \over 1 - S_0
  \sqrt{1 - \Gamma}} S_0 \sqrt{\Gamma}, \label{eq:tunS}
\end{equation}
where $\Gamma$ is a diagonal $N \times N$ matrix with
$\Gamma_{jj} = 0$ for index $j$ corresponding to the
current-carrying leads $1$ and $2$ and $\Gamma_{jj} = \Gamma_3$
for index $j$ corresponding to the voltage probe. The distribution
of $S_0$ and its derivatives is as described in Eqs.\
(\ref{eq:param})--(\ref{eq:dist}) at the beginning of
this section.
We can then find the distribution of $S$ from Eq.\ (\ref{eq:tunS})
and the distribution of its derivatives upon differentiating
Eq.\ (\ref{eq:tunS}).

\begin{figure}
  \epsfxsize=0.9\hsize
  \hspace{0.03\hsize}
  \epsffile{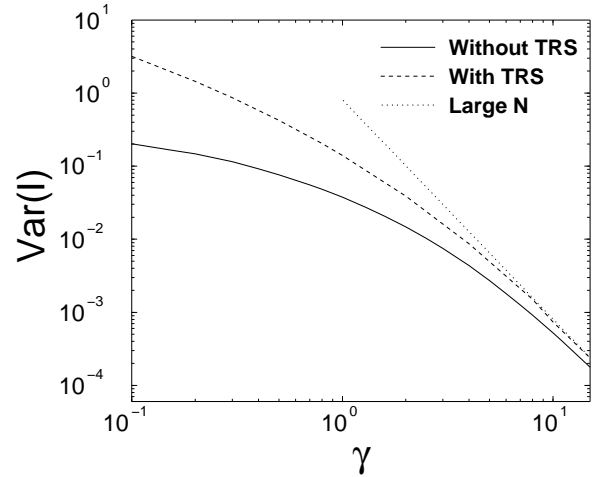}\\

\caption{ \label{fig:5} The variance of the pumped current in the case of
intrinsic dephasing as a function of
dimensionless dephasing rate $\gamma$ in the presence (dashed curve) and 
absence (solid curve) of time-reversal symmetry. The asymptotic result 
(dotted curve) is also shown. There is one open channel in each lead
($N_1=N_2=1$).}

\end{figure}

\begin{figure}
  \epsfxsize=0.9\hsize
  \hspace{0.03\hsize}
  \epsffile{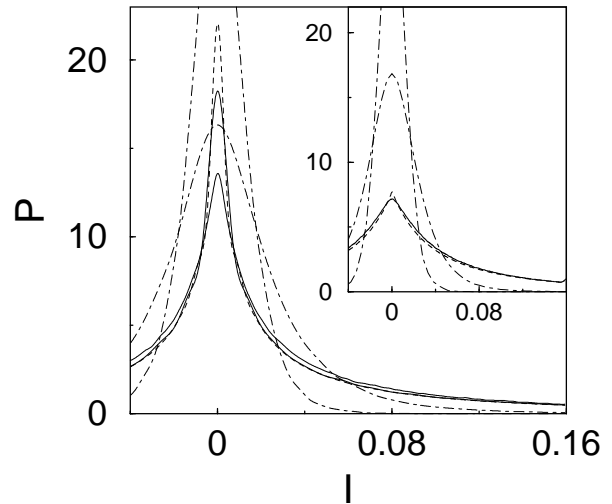}\\

\caption{ \label{fig:6} The probability distribution of the pumped current
with $N_1=N_2=1$ in the presence of time-reversal symmetry and
intrinsic dephasing. The dashed curve is the probability distribution in the
absence of dephasing. The two solid curves have dimensionless dephasing rates
$\gamma=.1$ (highest) and $\gamma=1$ (lowest). The dot-dashed curves have
dephasing rates $\gamma=8$ (lowest) and $\gamma=15$ (highest). The inset
shows the probability distribution of the pumped current
with $N_1=N_2=1$ in the absence of time-reversal symmetry with the same
parameters. The distribution with $\gamma=.1$ is indistinguishable from the case
without dephasing ($\gamma=0$).
}
\end{figure}

We have calculated the variance $\langle i^2 \rangle$ and the
full distribution of the current for $N_1 = N_2 = 1$
using Monte-Carlo integration with the above distribution. The
results are shown in Figs.\ \ref{fig:5} and \ref{fig:6}. We note
that intrinsic dephasing cuts off the tails of the distribution
$P(i)$. This is in contrast to the case of controlled
dephasing by a few-channel voltage
probe with ballistic point contacts, which merely replaces the
algebraic tail of the distribution at zero dephasing
by another, faster decaying, algebraic tail. As for controlled
dephasing,  the variance of the pumped
current is larger with time-reversal symmetry than without, though
for intrinsic dephasing
the difference is smaller than in the case of controlled dephasing
(cf.\ Fig.\ \ref{fig:2}). At large dephasing rates, the dependence
of $\langle i^2 \rangle$ on the presence or absence of time-reversal
symmetry vanishes. Also note that the probability to find
zero pumped current initially decreases when the dephasing rate is
increased, although the effect is not as strong as in the case of
controlled dephasing. For large dephasing rates, both intrinsic
and controlled dephasing yield the same distribution $P(i)$.

For large $N_1$ and $N_2$ the distribution $P(i)$ is again Gaussian,
with zero mean and with variance given by Eq.\ (\ref{eq:asym}) above
with $N_3$ replaced by $\gamma$. This result agrees with what was 
previously obtained
by Shutenko {\em et al.} using a different method.\cite{SAA}

\section{Conclusion}
\label{sec4}
In summary, we have derived the current distribution of electrons pumped 
adiabatically through a chaotic quantum dot, in the presence of a
voltage probe. The voltage can either serve as a controlled source of
dephasing, or as an effective description of intrinsic dephasing
processes inside the quantum dot.
For the case of a quantum dot with two single-mode current-carrying
point contacts, the conductance distribution is non-Gaussian with
algebraic tails. 
Remarkably, while dephasing shifts the weight of the probability
distribution $P(I)$ of the pumped current towards zero current,
a small amount of dephasing actually reduces the probability $P(I=0)$ to 
find no pumped current at all. This can be understood if the
probability to find zero pumped current is enhanced by destructive 
interference --- note the cusp in $P(I)$ at $I=0$ ---, which is
then suppressed by dephasing.
For a quantum dot with many-channel
point contacts (as well as for a quantum dot with single-channel
point contacts and a large dephasing rate), 
the current distribution is Gaussian. The width
of the distribution decreases monotonically with increasing dephasing
rate.
For dephasing rates $\gamma_{\phi}$ much larger than the escape
rate $\gamma \sim (N_1 + N_2)\Delta$ to the reservoirs, the r.m.s.\ 
current decays $\propto (\gamma/\gamma_{\phi})^{3/2}$ (see also
Ref.\ \onlinecite{SAA}). 

Not only dephasing, but also thermal smearing can reduce the size of
the pumped current. Thermal smearing has been considered by Shutenko
{\em et al.},\cite{SAA} who found that, without dephasing, the r.m.s.\
current decays $\propto (\gamma/T)^{-1/2}$ for temperature
$T \gg \gamma$. Experimentally, the dephasing rate $\gamma_{\phi}$ has
been found to increase $\propto T$ (or faster), which implies that,
for such a dephasing rate, dephasing is the more effective
mechanism to reduce the current pumped in a quantum pump.

It is interesting to compare the quantum pump to a (quantum)
rectifier. The latter device generates a dc current in response
to an a.c.\ bias voltage. For a setup similar to the quantum dot
pump considered here, the variance of the rectified current is 
proportional to $(\partial G/\partial X)^2$.\cite{rectif} 
For large dephasing rates, typically $(\partial G/\partial X)^2 \sim
(\gamma/\gamma_{\phi})^3$,\cite{BReview} 
so that the r.m.s.\ rectified current decays 
$\propto (\gamma/\gamma_{\phi})^{3/2}$.
As in an experimental realization a true quantum pump may coexist
with a rectifier (the a.c.\ voltages arising from displacement 
currents and parasitic capacitive coupling),\cite{switkesthesis} 
we conclude that
dephasing does not change the relative importance of one mechanism
over the other.

\acknowledgements

We thank B.I.\ Halperin and C. Marcus for stimulating discussions. 
Shortly before completion of our manuscript, we learned of Ref.\
\onlinecite{MB}, where the effect of dephasing in an electron 
pump is studied with an approach similar to ours.
This work was supported by the
NSF under grant nos.\ DMR 0086509 and DMR 9981283
and by the Sloan foundation.

\appendix
\section{}\label{appA}

In this appendix we outline the calculation of the moments
$\langle i^2 \rangle$ in the presence of controlled dephasing. A
detailed report of the calculations will appear in Ref.\
\onlinecite{cremers}.

In order to calculate the average $\langle i^2 \rangle$, we need
to perform an average over the $N \times N$ 
matrices $U$, $U'$, $Q_1$, and $Q_2$
defined in Eq.\ (\ref{eq:param}). The matrices $U$ and $U'$ are uniformly 
distributed in the unitary group [$U' = U^{\rm T}$ in the presence
of time-reversal symmetry (TRS)], while the distribution of the matrices
$Q_1$ and $Q_2$ is given by Eq.\ (\ref{eq:dist}). First we perform
the average over $Q_1$ and $Q_2$. Hereto, we parameterize $Q_1$ and
$Q_2$ as
\begin{equation}
  Q_{i}=\Psi^{\dagger-1}H_{i}\Psi^{-1},\ \ i=1,2,\label{eq:qham}
\end{equation}
where $H_1$ and $H_2$ are hermitian $N \times N$
matrices (real symmetric in the
presence of time-reversal symmetry) and $\Psi$ is a complex (real)
$N \times N$ matrix such that\cite{footpsi}
\begin{equation}
Q_{\varepsilon}= \Psi^{\dagger-1}\Psi^{-1}. \label{eq:Qpsi}
\end{equation}
Substitution of Eqs.\ (\ref{eq:qham}) and (\ref{eq:Qpsi}) into 
the distribution (\ref{eq:dist}) shows that the elements of $H_1$
and $H_2$ have a Gaussian distribution with zero mean and with
variance
\begin{equation}
\left\langle H_{ij}H_{kl}\right\rangle =\left\{ \begin{array}{ll} 
 4\left(\delta _{il}\delta _{jk}+ \delta _{ik}\delta _{jl}\right)& 
\mbox{TRS}, \\
 4\delta _{il}\delta _{jk}  & \mbox{no TRS}. \end{array} \right.
  \label{eq:Hvar} 
\end{equation}

We first perform the Gaussian average over the matrices $H_1$ and
$H_2$. The resulting expression contains the matrix $\Psi$
in the combination $Q_{\varepsilon}=\Psi^{\dagger-1}\Psi^{-1}$
only. We decompose $Q_{\varepsilon}$
in terms of its eigenvalue matrix $\hat\tau$ and a random unitary
(orthogonal) matrix $V$, 
\begin{equation}
  Q_{\varepsilon}=V \hat\tau V^{\dagger}.\label{eq:uu}
\end{equation}
and average over $V$ using the diagrammatic method of Ref.\ 
\onlinecite{B4} (or its straightforward generalization to orthogonal
matrices $V$ in the case when time-reversal symmetry is present).
The result of the average is an expression that depends on the
scattering matrix $S$ and the matrix of eigenvalues $\hat\tau$ only.

In the absence of time-reversal symmetry, the result of this calculation is
\begin{eqnarray}\nonumber
\left\langle i_{\rm dir}^2\right\rangle&=&
{8 \over \pi^2 N (N^2-1)} 
  \left( \mbox{tr}\, \hat\tau^2 (\mbox{tr}\, \hat\tau)^2 - 
    \mbox{tr}\, \hat\tau^4 \right) 
  \nonumber \\ && \mbox{} \times
  \left( N_1 N_2 + N_3 {N_1 G_{23}^2 +
    N_2 G_{13}^2 \over (G_{13} + G_{23})^2} \right)
\label{eq:fpumpa}
\end{eqnarray}
while in the presence of time-reversal symmetry, we find
\begin{eqnarray}
  \left\langle i_{\rm dir}^2\right\rangle&=&
  {8 \over \pi^2 N (N-1)}
  \left( G_{12} + {G_{13} G_{23} \over G_{13} + G_{23}} \right)
  \left( \mbox{tr}\, \hat\tau^2 (\mbox{tr}\, \hat\tau)^2
   \right. \nonumber \\ && \left. \mbox{} 
  - (\mbox{tr}\, \hat\tau^2)^2 +
  2 \mbox{tr}\, \hat\tau (\mbox{tr}\, \hat\tau^3) -
  2 \mbox{tr}\, \hat\tau^4 \right).
\end{eqnarray}
The contribution from
$i_{\rm rect}$ is a factor $N$
smaller than $i_{\rm dir}$ for large $N$. Now, the large-$N$
result, Eq.\ (\ref{eq:asym}),
follows easily employing the known density of dimensionless delay
times,
\begin{eqnarray}
\rho(\tau) &=& \left\langle \sum_{j} \delta(\tau-\tau_j) \right\rangle
  \nonumber \\ &=&
\frac{N}{2\pi\tau^2}\label{eq:dens}
\sqrt{(\tau_+-\tau)(\tau-\tau_-)},
\end{eqnarray}
where
$\tau_\pm=(3\pm \sqrt{8})/N$, to integrate over $\hat\tau$.

For the case $N_1 = N_2 = 1$, we were able to obtain exact expressions for
$\left\langle i_{\rm dir}i_{\rm rect}\right\rangle$ and
$\left\langle i_{\rm rect}^{2}\right\rangle$ for arbitrary $N_3$
and time-reversal symmetry is broken. These expressions, which
contain up to a product of four traces involving the matrix
$\hat \tau$, were too lengthy to be reported here, and will be
published in Ref.\ \onlinecite{cremers}.
The results for $\langle i^2 \rangle$ are
shown in Fig.\ \ref{fig:2}. We summarize the main steps of the
calculation below. 
Starting from Eq.\ (\ref{eq:fpumpa}) and similar expressions 
for the variance of $i_{\rm rect}$,\cite{cremers} 
we integrate over the  eigenvalues $\tau_j$ of
the dimensionless time-delay matrix $Q_{\varepsilon}$ and the 
scattering matrix $S$. To average over the $\tau_j$,
$j=1,\ldots,N$, we introduce the
dimensionless escape rates $x_n=1/\tau_n$ which are
distributed according to the generalized Laguerre ensemble,\cite{B3}
\begin{equation}
P(x_1, \ldots, x_N)\propto \prod_{n<m} \left(x_n-x_m\right)^2\ \prod_n\, x_n^N\,
\mbox{exp}(x_n).\label{eq:probtau}
\end{equation}
Since the $\tau_n$ appear only in products of up to four traces, we
need the marginal $n$-point distributions $R_n(x_1,\ldots,x_n)$ for
$n=1,\ldots,4$ only, where
\begin{eqnarray*}
R_n(x_1,\ldots,x_n) &=&\frac{N!}{(N-n)!}
  \\ && \mbox{} \times \int_{0}^{\infty}dx_{n+1}\ldots dx_N
P(x_1,\ldots,x_N).
\end{eqnarray*}
We can find exact expressions of $R_1,\ldots,R_4$ using the associated Laguerre
polynomials $L^N_n(x)$ which are orthogonal with respect to the weight function
$x^N e^{-x}$. in particular, for the case of broken time-reversal symmetry,
$$
R_n(x_1,\ldots,x_n)=\mbox{det}\left[K(x_i,x_j)\right]_{i,j=1,\ldots,n},
$$
where
$$
K(x_i,x_j)=\sum_{l=1}^N L^N_l(x_i) L^N_l(x_j) e^{(x_i+x_j)/2}
\left(x_ix_j\right)^{N/2}.
$$
The average over $S$ can be done using the polar 
decomposition of $S$,
$$
S=
\left(\begin{array}{ll}
u&0\\
0&v
\end{array}\right)
\left(\begin{array}{cc}
\sqrt{1-t^{\dagger}t}&it^{\dagger}\\
it&\sqrt{1-t^{\dagger}t}
\end{array}\right)
\left(\begin{array}{ll}
u'&0\\
0&v'
\end{array}\right),
$$
where $u$ and $u'$ ($v$ and $v'$) are $2\times2$ $(N_3\times N_3)$ 
unitary matrices
and $t$ is an $N_3\times 2$ matrix with all elements equal to zero except
$t_{nn}=\sqrt{T_n}, n=1,2$. In the presence of time-reversal symmetry,
$u' = u^{\rm T}$ and $v' = v^{\rm T}$.
The parameters $T_1$ and $T_2$ govern the escape rate into the voltage 
probe. The uniform distribution of $U$ and $U'$ in the unitary group
yields the integration measure
$$
dS=\left|T_1-T_2\right|^2 \left(T_1T_2\right)^{N_3-2}dudu'dvdv'dT_1dT_2.
$$
The average over $S$ then
reduces to an integral over $T_1,T_2$ and one angle 
$\phi$ uniformly distributed in the interval $0 < \phi < \pi/2$.
In terms of these variables we can write
\begin{eqnarray}
G_{13}&=&T_1 \mbox{cos}^2 \phi+T_2
\mbox{sin}^2 \phi\nonumber\\
G_{23}&=&T_1 \mbox{sin}^2 \phi+T_2
\mbox{cos}^2 \phi\nonumber\\
\mbox{tr}\,S_{33}^{\phantom{\dagger}}S_{33}^{\dagger}
S_{33}^{\phantom{\dagger}}S_{33}^{\dagger}&=&N-4+(1-T_1)^2+(1-T_2)^2
\nonumber\\
\mbox{tr}\,S_{13}^{\dagger}S_{13}^{\phantom{\dagger}}S_{33}^{\dagger}
S_{33}^{\phantom{\dagger}}&=&(1-T_1)T_1\mbox{cos}^2\phi
+(1-T_2)T_2\mbox{sin}^2\phi\nonumber,
\end{eqnarray}
where $S_{ij}$ denotes the $ij$ block in the scattering matrix $S$. (The last two
terms appear in the expression for $i_{\rm rect}$.\cite{cremers})

\section{}\label{appB}

In order to obtain the full distribution of the pumped current $i$, we 
numerically generated matrices $U$, $U'$, $Q_1$, and $Q_2$ according
to the appropriate distributions. The matrices $U$ and $U'$ are 
uniformly distributed in the unitary group ($U' = U^{\rm T}$ in the
presence of time-reversal symmetry). Using a parameterization
in terms of Euler angles,\cite{Kus} their generation is relatively
straightforward. The numerical generation of matrices $Q_1$ and $Q_2$
according to the distribution (\ref{eq:dist})
makes use of a trick that was inspired by Ref.\ \onlinecite{Brezin},
which would like to explain below.
 
We parameterize the $N \times N$ matrices $Q_1$, $Q_2$, and 
$Q_{\varepsilon}$ as
\begin{eqnarray}
  Q_{\varepsilon}&=&C^{{\rm T}-1}C^{-1}, \nonumber \\
  Q_i &=& C^{{\rm T}-1} H_i C^{-1},\ \ i=1,2 \label{eq:QQi}
\end{eqnarray}
where $C$ is a complex $N \times 2N$ matrix 
[real $N \times (2N+1)$ matrix in
the presence of time-reversal symmetry] and $H_i$ ($i=1,2$) is a
hermitian $2N \times 2N$ matrix [real symmetric $(2N+1) \times (2N+1)$
matrix in the presence of time-reversal symmetry]. In Eq.\
(\ref{eq:QQi}), the inverse $C^{-1}$ is defined as the right-inverse,
$C^{-1} = C^{\rm T} (C C^{\rm T})^{-1}$. In Ref.\
\onlinecite{Brezin} it was shown that the matrix $Q_{\varepsilon}$ has
the distribution (\ref{eq:dist}) if the elements of $C$ are all chosen
from a Gaussian distribution with unit variance. From there, one
can show by substitution of Eq.\ (\ref{eq:QQi}) into Eq.\ (\ref{eq:dist}),
that the parameterization (\ref{eq:QQi}) 
reproduces the correct distribution (\ref{eq:dist}) for all three
matrices $Q_{\varepsilon}$, $Q_1$, and $Q_2$ if the elements of
the matrices $H_1$ and $H_2$ are all chosen from a Gaussian
distribution with variance given by Eq. (\ref{eq:Hvar}) above.


\begin{references}
\bibitem{LesHouches}
      {\em Mesoscopic Quantum
      Physics}, edited by E. Akkermans, G. Montambaux,
      J.-L. Pichard, and J. Zinn-Justin (North-Holland, Amsterdam,
      1995).
\bibitem{Kouwenhoven} 
      L. P. Kouwenhoven, C. M. Marcus, P. L. McEuen, 
      S. Tarucha, R. M. Westervelt, and N. S. Wingreen, 
      in {\em 
      Mesoscopic Electron Transport}, edited by L. L. Sohn, L. P. 
      Kouwenhoven, and G. Sch\"on (Kluwer, Dordrecht, 1997).

\bibitem{switkes} 
      M. Switkes, C. M. Marcus, K. Campman,
      and A. C. Gossard, Science {\bf 283}, 1905 (1999).

\bibitem{SZB} 
      B. Spivak, F. Zhou, and M. T. Beal Monod, Phys. Rev. B
      {\bf 51}, 13226  (1995).

\bibitem{B1} P. W. Brouwer, Phys. Rev. B {\bf 58}, 10135 (1998).

\bibitem{ZSA} 
      F. Zhou, B. Spivak, and B. L. Altshuler, Phys. Rev. Lett.
      {\bf 82}, 608 (1999).

\bibitem{AG} 
      B. L. Altshuler and L. I. Glazman, Science {\bf 283}, 1864 (1999).

\bibitem{SAA} 
      T. A. Shutenko, I. L. Aleiner, and B. L. Altshuler,
      Phys. Rev. B {\bf 61}, 10366 (2000).

\bibitem{Avron}
       J. E. Avron, A. Elgart, G. M. Graf, and L. Sadun, 
       Phys. Rev. B {\bf 62}, 10618 (2000).

\bibitem{VAA}
      M. G. Vavilov, V. Ambegaokar and I. L. Aleiner, 
      Phys. Rev. B {\bf 63}, 195313 (2001).

\bibitem{AAK}
I. L. Aleiner, B. L. Altshuler, and A. Kamenev,
  Phys. Rev. B 62, 10373 (2000).

\bibitem{polianski} M. L. Polianski and P. W. Brouwer, Phys. Rev. B
{\bf 64}, 75304 (2001).

\bibitem{KJVHF} 
      L. P. Kouwenhoven, A. T. Johnson, N. C. van der Vaart,
      C. J. P. M. Harmans, and C. T. Foxon, 
       Phys. Rev. Lett. {\bf 67}, 1626 (1991).
\bibitem{PLUED} 
      H. Pothier, P. Lafarge, C. Urbina, D. Esteve, and M. H. Devoret, 
      Europhys. Lett.\ {\bf 17}, 249 (1992).

\bibitem{but3}M. B\"uttiker, Phys. Rev. B {\bf 33}, 3020 (1986); IBM J. Res.
Dev. {\bf 32}, 63 (1988).
\bibitem{BM1}
H. U. Baranger and P. A. Mello, Phys. Rev. B {\bf 51}, 4703 (1995).
\bibitem{Brouwer}P. W. Brouwer and C. W. J. Beenakker, Phys. Rev. B {\bf 55},
4695 (1997).
\bibitem{MB} M. Moskalets and M. B\"uttiker, cond-mat/0108061.

\bibitem{ABG} I. L. Aleiner, P. W. Brouwer, and L. I. Glazman,
cond-mat/0103008.

\bibitem{but2} M. B\"uttiker, J. Phys. Condens. Matter {\bf 5},
9361 (1993); M. B\"uttiker, H. Thomas, and A. Pr\^etre, Z. Phys. B {\bf 94},
133 (1994).
\bibitem{B3} P. W. Brouwer, K. M. Frahm, and C. W. J. Beenakker,
  Phys. Rev. Lett. {\bf 78}, 4737 (1997);
{\it Waves in Random Media} {\bf 9}, 91 (1999).
\bibitem{B4}P. W. Brouwer and C. W. J. Beenakker, J. Math. Phys. {\bf 37},
4904 (1996).
\bibitem{MelloSeligman}P. A. Mello, P. Pereyra and T. H. Seligman, 
Ann. Phys. {\bf 161}, 254 (1985).
\bibitem{B5}P. W. Brouwer, Phys. Rev. B {\bf 51}, 16878 (1995).


\bibitem{rectif} P. W. Brouwer, Phys. Rev. B {\bf 63}, 121303 (2001).
\bibitem{BReview} C. W. J. Beenakker, Rev. Mod. Phys.
{\bf 69}, 731 (1997).
\bibitem{switkesthesis} M. Switkes, Ph. D. thesis (Stanford, 1999).

\bibitem{footpsi}
Note that Eq.\ (\ref{eq:Qpsi}) only fixes $\Psi$ up to
right-multiplication with a unitary matrix (orthogonal matrix in 
the presence of time-reversal symmetry). However, this unitary
(orthogonal) matrix can be absorbed into $H_1$ and $H_2$, the
distribution of which is invariant under unitary (orthogonal)
transformations.
\bibitem{cremers} J.N.H.J. Cremers, Ph.D. Thesis, Harvard University,
  in preparation.
\bibitem{Kus} K. Zyczkowski and M. Kus, Phys. Rev. E
{\bf 53}, 319 (1996).

\bibitem{Brezin}E. Br\'{e}zin, S. Hikami, and A. Zee, Nucl. Phys. B
{\bf 464}, 411 (1996).

\end{references}
\end{document}